\documentclass[12pt]{article}\usepackage{deflat-id}
\input{MyQcircuit.sty}
\begin{document}

\title{Replacing Two Controlled-U's \\
with Two CNOTs}

\author{Robert R. Tucci\\
        P.O. Box 226\\
        Bedford,  MA   01730\\
        tucci@ar-tiste.com}

\date{ \today}

\maketitle

\vskip2cm
\section*{Abstract}
We give a simple,
closed-form formula,
what we call the Deflation Identity,
for converting any 2-qubit circuit with
exactly two controlled-U's
(and some 1-qubit rotations)
into an equivalent circuit
with just two CNOTs (and some 1-qubit rotations).
We also give two interesting applications
of the Deflation Identity; one to ``opening
and closing a breach" in a quantum
circuit, the other to
the CS decomposition of
a 2-qubit operator.

\section{Introduction and Motivation}

In 1995, Ref.\cite{Bar95} showed
that any controlled-U gate can be
expressed with just 2 CNOTs
(and some 1-qubit rotations); i.e.,

\beq
\begin{array}{c}
\Qcircuit @C=1em @R=1em @!R{
&\emptygate
&\qw
\\
&\dotgate\qwx[-1]
&\qw
}
\end{array}
=
\begin{array}{c}
\Qcircuit @C=1em @R=1em @!R{
&\emptygate
&\timesgate
&\emptygate
&\timesgate
&\emptygate
&\qw
\\
&\qw
&\dotgate\qwx[-1]
&\qw
&\dotgate\qwx[-1]
&\qw
&\qw
}
\end{array}
\;,
\eeq
where the empty boxes represent
2-dimensional unitary matrices.

In 2003,
Vidal and Dawson\cite{VD} showed that
any 2-qubit operator can be
expressed by a circuit
containing just 3 CNOTs
(and some 1-qubit rotations).

In light of the Vidal-Dawson
result, a quantum circuit that
contains exactly two controlled-U's
does not require 4 CNOTs
(i.e., twice
the number of CNOTs required by one controlled-U)
to express; rather, it can be
expressed with just 3 CNOTs.
Actually, it turns out that
even 3 CNOTs is more than what is
needed. In  this paper
we show that
any quantum circuit with
exactly two controlled-U's
can be
expressed with just 2 CNOTs;
i.e., we show that

\begin{subequations}\label{eq-emblem}
\beq
\begin{array}{c}
\Qcircuit @C=1em @R=1em @!R{
&\emptygate
&\emptygate
&\emptygate
&\qw
\\
&\dotgate\qwx[-1]
&\emptygate
&\dotgate\qwx[-1]
&\qw
}
\end{array}
=
\begin{array}{c}
\Qcircuit @C=1em @R=1em @!R{
&\emptygate
&\timesgate
&\emptygate
&\timesgate
&\emptygate
&\qw
\\
&\emptygate
&\dotgate\qwx[-1]
&\emptygate
&\dotgate\qwx[-1]
&\emptygate
&\qw
}
\end{array}
\;,
\eeq
and

\beq
\begin{array}{c}
\Qcircuit @C=1em @R=1em @!R{
&\emptygate
&\emptygate
&\dotgate\qwx[1]
&\qw
\\
&\dotgate\qwx[-1]
&\emptygate
&\emptygate
&\qw
}
\end{array}
=
\begin{array}{c}
\Qcircuit @C=1em @R=1em @!R{
&\emptygate
&\timesgate
&\emptygate
&\timesgate
&\emptygate
&\qw
\\
&\emptygate
&\dotgate\qwx[-1]
&\emptygate
&\dotgate\qwx[-1]
&\emptygate
&\qw
}
\end{array}
\;.
\eeq
\end{subequations}

In this paper, we give a simple,
closed-form formula,
what we call the Deflation Identity,
for performing the
circuit conversions
illustrated by Eq.(\ref{eq-emblem}).
We also give two interesting applications
of the Deflation Identity; one to ``opening
and closing a breach" in a quantum
circuit, the other to
the CS decomposition of
a 2-qubit operator.

Note that I am including with the
paper an Octave/Matlab m-file
that checks the Deflation Identity numerically.

\section{Notation}
In this section, we will
define some notation that is
used throughout this paper.
For additional information about our
notation, see Ref.\cite{Paulinesia}.

As usual, $\RR, \CC$
will stand for the real and complex numbers,
respectively.
For any complex matrix $A$, the symbols
$A^*, A^T, A^\dagger$ will stand
for the complex conjugate, transpose,
and Hermitian conjugate, respectively,
of $A$.
(Hermitian conjugate
a.k.a. conjugate transpose and
adjoint)

The Pauli matrices are defined by:

\beq
\sigx=
\left(
\begin{array}{cc}
0 & 1\\
1 & 0
\end{array}
\right)
\;,
\;\;
\sigy=
\left(
\begin{array}{cc}
0 & -i\\
i & 0
\end{array}
\right)
\;,
\;\;
\sigz=
\left(
\begin{array}{cc}
1 & 0\\
0 & -1
\end{array}
\right)
\;.
\eeq
They satisfy
\beq
\sigx \sigy = -\sigy \sigx = i \sigz
\;,
\eeq
and the two other equations
obtained from this one by permuting
the indices $(X,Y,Z)$ cyclically.
We will also have occasion
to use the operators $n=\frac{1}{2}(1-\sigz)$,
$\nbar = 1-n$ and $\vec{\sigma}= (\sigx, \sigy, \sigz)$.
For any $\vec{\theta}, \vec{a}\in \RR^3$,

\beq
e^{-i\frac{\vec{\theta}}{2}\cdot\vec{\sigma}}
(\vec{\sigma}\cdot\vec{a})
e^{ i\frac{\vec{\theta}}{2}\cdot\vec{\sigma}}
=
\vec{\sigma}\cdot\vec{b}
\;,
\eeq
where $\vec{b}$ is the vector obtained
by rotating $\vec{a}$ by an angle $|\vec{\theta}|$,
with respect to an axis
parallel to $\vec{\theta}$.
(The sense of the rotation is determined by the right
hand rule, meaning that if the thumb of your right
hand points in the direction
$\vec{\theta}$, then your other
right hand fingers point from $\vec{a}$ to $\vec{b}$.)

In our quantum circuits,
$\begin{array}{c}
\Qcircuit @C=1em @R=.5em @!R{&\emptygate &\qw}
\end{array}$ will represent a
2 dimensional unitary matrix, whereas
$\begin{array}{c}
\Qcircuit @C=1em @R=.5em @!R{&\gate{R_D} &\qw}
\end{array}$ with $D=X,Y,Z$ will represent
a 1-qubit rotation $e^{i\theta\sigma_D}$
in the $D$ direction.

Let $\sigma_{X_\mu}$
for $\mu \in\{0,1,2,3\}$
be defined by
$\sigma_{X_0}=\sigma_1 = I_2$,
where $I_2$ is the 2 dimensional identity matrix,
$\sigma_{X_1}=\sigx$,
$\sigma_{X_2}=\sigy$,
and
$\sigma_{X_3}=\sigz$.
Now define

\beq
\sigma_{X_\mu X_\nu}=
\sigma_{X_\mu}\otimes \sigma_{X_\nu}
\;
\eeq
for $\mu, \nu\in\{0,1,2,3\}$.
For example, $\sigxy= \sigx\otimes \sigy$
and $\sigux = I_2\otimes\sigx$.
The matrices $\sigma_{X_\mu X_\nu}$
satisfy
\beq
\sigxx \sigyy = \sigyy \sigxx = -\sigzz
\;,
\label{eq-com-sigxx-sigyy}
\eeq
and
the two other equations
obtained from this one by permuting
the indices $(X,Y,Z)$ cyclically.

We will also have occasion to use the
exchange operator for two qubits,
defined by

\beq
\udarrow =
\left[
\begin{array}{cccc}
1 & 0 & 0 & 0\\
0 & 0 & 1 & 0\\
0 & 1 & 0 & 0\\
0 & 0 & 0 & 1\\
\end{array}
\right]
\;.
\eeq
(If the rows and columns
of this matrix are labelled by
$(a_1, a_0)=00, 01, 10, 11$,
where $a_0$ labels the state of qubit $0$
and $a_1$ that of qubit $1$,
then
$\udarrow$ indeed exchanges the two qubits.)
A useful identity involving $\udarrow$ is:

\beq
\left[
\begin{array}{cccc}
a & 0   & b &  0\\
 0  & \alpha & 0   & \beta\\
c & 0   & d &  0\\
 0  & \gamma &  0  & \delta\\
\end{array}
\right]
=
\udarrow
\left[
\begin{array}{cccc}
a & b & 0 & 0\\
c & d & 0 & 0\\
 0  & 0   & \alpha & \beta\\
 0  & 0   & \gamma & \delta\\
\end{array}
\right]
\udarrow
\;.
\eeq

Define

\beq
\unor=
\frac{1}{\sqrt{2}}
\left[
\begin{array}{cccc}
 1 & 0 & 0 & i\\
 0 & i & 1 & 0\\
 0 & i &-1 & 0\\
 1 & 0 & 0 &-i
\end{array}
\right]
\;.
\eeq
It is easy to check that
$\unor$ is a unitary matrix.
The columns of
$\unor$ are an orthonormal basis,
often called the ``magic basis" in the
quantum computing literature. (That's
why we have chosen to
call this matrix $\unor$,
because of the ``m" in magic).

In this paper, we often
need to find the outcome of
applying a similarity transformation
(i.e., a change of
basis)
$\unor^\dagger(\cdot)\unor$ (or
$\unor(\cdot)\unor^\dagger$)
to a matrix $X\in \CC^{4\times 4}$. Since $X$ can
always be expressed as a linear
combination of the $\sigma_{X_\mu X_\nu}$,
it is useful to know
the outcomes
$\unor^\dagger(\sigma_{X_\mu X_\nu})\unor$
(or $\unor(\sigma_{X_\mu X_\nu})\unor^\dagger$)
for $\mu, \nu\in\{0,1,2,3\}$. One finds
the following two tables:

\beq
\unor^\dagger(A\otimes B)\unor=\;\;\;\;
\begin{array}{|rr|rrrr|}\hline
&&& B &\rarrow&\\
            &       &1       & \sigx  & \sigy  & \sigz\\ \hline
            &1      &1       &-\siguy & \sigyz &-\sigyx\\
A           &\sigx  &-\sigzy & \sigzu &-\sigxx &-\sigxz\\
\downarrow  &\sigy  &-\sigyu & \sigyy &-\siguz & \sigux\\
            &\sigz  &-\sigxy & \sigxu & \sigzx & \sigzz\\\hline
\end{array}
\;,\label{tab-mh-m}
\eeq
and

\beq
\unor(A\otimes B)\unor^\dagger=\;\;\;\;
\begin{array}{|rr|rrrr|}\hline
&&& B &\rarrow&\\
            &       &1       & \sigx  & \sigy  & \sigz\\ \hline
            &1      &1       & \sigyz &-\sigux &-\sigyy\\
A           &\sigx  & \sigzx &-\sigxy &-\sigzu &-\sigxz\\
\downarrow  &\sigy  &-\sigyu &-\siguz & \sigyx & \siguy\\
            &\sigz  & \sigxx & \sigzy &-\sigxu & \sigzz\\\hline
\end{array}
\;.\label{tab-m-mh}
\eeq

Another similarity transformation that
we shall often encounter in his paper is
$\uutoz(\cdot)\uutoz$ (and its
twin $\zztou(\cdot)\zztou$). One finds
\beq
\uutoz(A\otimes B)\uutoz=\;\;\;\;
\begin{array}{|rr|rrrr|}\hline
&&& B &\rarrow&\\
            &       &1       & \sigx  & \sigy  & \sigz\\ \hline
            &1      &1       & \sigux & \sigzy & \sigzz\\
A           &\sigx  & \sigxx & \sigxu & \sigyz &-\sigyy\\
\downarrow  &\sigy  & \sigyx & \sigyu &-\sigxz & \sigxy\\
            &\sigz  & \sigzu & \sigzx & \siguy & \siguz\\\hline
\end{array}
\;,\label{tab-uutoz}
\eeq
and

\beq
\zztou(A\otimes B)\zztou=\;\;\;\;
\begin{array}{|rr|rrrr|}\hline
&&& B &\rarrow&\\
            &       &1       & \sigx  & \sigy  & \sigz\\ \hline
            &1      &1       & \sigxx & \sigxy & \siguz\\
A           &\sigx  & \sigxu & \sigux & \siguy & \sigxz\\
\downarrow  &\sigy  & \sigyz & \sigzy &-\sigzx & \sigyu\\
            &\sigz  & \sigzz &-\sigyy & \sigyx & \sigzu\\\hline
\end{array}
\;.\label{tab-zztou}
\eeq

\section{Statement and Proof of Deflation Identity}

In this section, we will present
Theorem \ref{th-defla1}, which
is referred to in this paper
as the Deflation Identity.
We will also present
a simple generalization,
given by Theorem \ref{th-defla2},
of the Deflation Identity .

\begin{theo}\label{th-defla1}
For any real numbers $\theta_L, \beta, \beta'$ and $\theta_R$, if
\beq
U=
\begin{array}{c}
\Qcircuit @C=1em @R=.5em @!R{
&\gate{e^{i\theta_L\sigz}}
&\gate{e^{i\beta\sigy}}
&\gate{e^{i\theta_R\sigz}}
&\qw
\\
&\dotgate\qwx[-1]
&\gate{e^{i\beta'\sigy}}
&\dotgate\qwx[-1]
&\qw
}
\end{array}
\;,
\label{eq-defla-u-in}
\eeq
then

\beq
U=
\begin{array}{c}
\Qcircuit @C=1em @R=.5em @!R{
&\gate{e^{i\frac{\theta_L}{2}\sigz}}
&\gate{e^{i\gamma_L\sigy}}
&\timesgate
&\gate{e^{i\mu\sigz}}
&\timesgate
&\gate{e^{i\gamma_R\sigy}}
&\gate{e^{i\frac{\theta_R}{2}\sigz}}
&\qw
\\
&\qw
&\gate{e^{i\gamma'_L\sigy}}
&\dotgate\qwx[-1]
&\gate{e^{i\mu'\sigx}}
&\dotgate\qwx[-1]
&\gate{e^{i\gamma'_R\sigy}}
&\qw
&\qw
}
\end{array}
\;,
\label{eq-defla-u-out}
\eeq
where the real numbers
$\gamma_L, \gamma_L', \mu, \mu', \gamma_R, \gamma_R'$
are defined as follows. Let

\beq
x_{1\pm} =
\cos(\frac{\theta_L + \theta_R}{2})
\cos(-\beta' \pm \beta)
\;,
\eeq

\beq
x_{2\pm} =
\cos(\frac{\theta_L - \theta_R}{2})
\sin(-\beta' \pm \beta)
\;,
\eeq
and

\beq
p_{\pm} = \sqrt{x_{1\pm}^2 + x_{2\pm}^2}
\;.
\eeq
Suppose $s\in \{+,-\}$.
If $p_{s}=0$, then
$\xi_{s}=0$. Otherwise,

\beq
\cos \xi_{\pm} = \frac{x_{1\pm}}{p_{\pm}}
\;,\;\;
\sin \xi_{\pm} = \frac{x_{2\pm}}{p_{\pm}}
\;.
\eeq
Let

\beq
y_{1\pm} =
\mp\sin(\frac{\theta_L - \theta_R}{2})
\sin(-\beta' \pm \beta)
\;,
\eeq

\beq
y_{2\pm} =
\mp\sin(\frac{\theta_L + \theta_R}{2})
\cos(-\beta' \pm \beta)
\;,
\eeq
and

\beq
q_{\pm} = \sqrt{y_{1\pm}^2 + y_{2\pm}^2}
\;.
\eeq
Suppose $s\in \{+,-\}$.
If $q_{s}=0$, then
$\eta_{s}=0$. Otherwise,

\beq
\cos \eta_{\pm} = \frac{y_{1\pm}}{q_{\pm}}
\;,\;\;
\sin \eta_{\pm} = \frac{y_{2\pm}}{q_{\pm}}
\;.
\eeq
The $\gamma$'s are defined by

\begin{subequations}
\beq
\gamma_L = \frac{1}{4}(
 \eta_{+}
-\eta_{-}
+\xi_{+}
-\xi_{-})
\;,
\eeq

\beq
\gamma'_L = \frac{1}{4}(
-\eta_{+}
-\eta_{-}
-\xi_{+}
-\xi_{-}
+\pi)
\;,
\eeq

\beq
\gamma_R = \frac{1}{4}(
-\eta_{+}
+\eta_{-}
+\xi_{+}
-\xi_{-})
\;,
\eeq
and

\beq
\gamma'_R = \frac{1}{4}(
+\eta_{+}
+\eta_{-}
-\xi_{+}
-\xi_{-}
-\pi)
\;.
\eeq
\end{subequations}
The $\mu$'s are defined by

\beq
\cos\mu_\pm = p_{\pm}
\;,\;\;
\sin\mu_\pm = q_{\pm}
\;,
\eeq
and

\beq
\mu  = \frac{\mu_+ - \mu_-}{2}
\;,\;\;
\mu' = \frac{\mu_+ + \mu_-}{2}
\;.
\eeq
\end{theo}
\proof

Note that

\beq
\begin{array}{c}
\Qcircuit @C=1em @R=.5em @!R{
&\gate{e^{i\theta\sigz}}
&\qw
\\
&\dotgate\qwx[-1]
&\qw
}
\end{array}
=
e^{i\theta\sigz(0)n(1)}=
e^{i\frac{\theta}{2}(\siguz - \sigzz)}
\;.
\eeq
For the matrix $U$ given by Eq.(\ref{eq-defla-u-in}),
we can define a matrix
$V$ by:

\beq
U=
e^{ i\frac{\theta_L}{2}\siguz}
V
e^{ i\frac{\theta_R}{2}\siguz}
\;.
\eeq
Thus, $V$ is merely $U$ with
the local $\siguz$ rotations
on both ends removed:

\beq
V=
e^{-i\frac{\theta_L}{2}\sigzz}
e^{ i(\beta'\sigyu + \beta\siguy)}
e^{-i\frac{\theta_R}{2}\sigzz}
\;.
\label{eq-v-tbt}
\eeq
Let

\beq
\tilde{V}
=
\unor^\dagger
V
\unor
\;.
\eeq
Applying the table given in
Eq.(\ref{tab-mh-m}) to
Eq.(\ref{eq-v-tbt}) yields:

\beq
\tilde{V}
=
e^{-i\frac{\theta_L}{2}\sigzz}
e^{ i(-\beta'\sigyu + \beta\sigyz)}
e^{-i\frac{\theta_R}{2}\sigzz}
\;.
\label{eq-til-v-tbt}
\eeq
We want to consider the real and
imaginary parts of $\tilde{V}$,
defined by:

\beq
\tilde{V}_1=
\frac{
\tilde{V}
+
(\tilde{V})^*
}{2}
\;,\;\;
\tilde{V}_2=
\frac{
\tilde{V}
-
(\tilde{V})^*
}{2i}
\;.
\label{eq-vr-vi-def}
\eeq
Using Eqs.(\ref{eq-til-v-tbt}) and
(\ref{eq-vr-vi-def}), we can find
more explicit expressions for $\tilde{V}_1$
and $\tilde{V}_2$.
If

\beq
C_L = \cos(\frac{\theta_L}{2})
\;,\;\;
S_L = \sin(\frac{\theta_L}{2})
\;,\;\;
\Lambda = -\beta'\siguy + \beta\sigzy
\;,
\eeq
these explicit expressions for $\tilde{V}_1$
and $\tilde{V}_2$ are:

\beq
\tilde{V}_1
=
\udarrow
[
C_L C_R e^{i\Lambda}
-
S_L S_R e^{-i\Lambda}
]
\udarrow
\;,
\eeq
and

\beq
\tilde{V}_2
=
\udarrow(-\sigzz)
[
S_L C_R e^{i\Lambda}
-
C_L S_R  e^{-i\Lambda}
]
\udarrow
\;.
\eeq
Now observe that
$\tilde{V}_1$
and $\tilde{V}_2$
are both of the following form:
for $j=1,2$,

\beq
\tilde{V}_{j}=
\udarrow
\left(
\begin{array}{cc}
A_{j+} & 0 \\
0 & A_{j-}
\end{array}
\right)
\udarrow
\;.
\label{eq-vj-aj}
\eeq

$A_{1\pm}$ is of the form

\beq
A_{1\pm} =
x_{1\pm} + i \sigy x_{2\pm}
=
p_{\pm} e^{i \sigy \xi_{\pm}}
\;.
\eeq
This last equation defines
$x_{1\pm}$, $x_{2\pm}$, $p_{\pm}$ and
$\xi_{\pm}$. If we define
$\xi_{\pm}^L$  and $\xi_{\pm}^R$ to be
arbitrary real numbers such that

\beq
\xi_{\pm}^L + \xi_{\pm}^R = \xi_{\pm}
\;,
\eeq
then

\beq
A_{1\pm}
=
e^{i\sigy\xi_{\pm}^L}
p_{\pm}
e^{i\sigy\xi_{\pm}^R}
\;.
\eeq

$A_{2\pm}$ is of the form

\beq
A_{2\pm}= y_{1\pm}\sigx + y_{2\pm}\sigz
= q_{\pm} e^{i\sigy\eta_{\pm}}\sigx
\;.
\eeq
This last equation defines
$y_{1\pm}$, $y_{2\pm}$, $q_{\pm}$ and
$\eta_{\pm}$. Define
$\phi_\pm$ by

\beq
\phi_\pm -\frac{\pi}{2}
=
-\eta_{\pm}
+\xi_{\pm}^L
-\xi_{\pm}^R
\;.
\eeq
Then

\beqa
e^{-i\sigy\xi_{\pm}^L}
\left(A_{2\pm}\right)
e^{-i\sigy\xi_{\pm}^R}&=&
e^{-i\sigy\xi_{\pm}^L}
\left(q_{\pm}e^{i\sigy\eta_{\pm}}\sigx\right)
e^{-i\sigy\xi_{\pm}^R}\\
&=&
q_{\pm}\sigx
e^{i\sigy(\phi_\pm-\frac{\pi}{2})}\\
&=&
q_{\pm}[\sin(\phi_\pm)\sigx
+
\cos(\phi_\pm)\sigz]\\
&=&
e^{-i\sigy \frac{\phi_\pm}{2}}
q_{\pm}\sigz
e^{i\sigy \frac{\phi_\pm}{2}}
\;.
\eeqa

Inserting these explicit expressions
for $A_{1\pm}$ and $A_{2\pm}$
into Eq.(\ref{eq-vj-aj})
yields

\beq
\tilde{V}=
\udarrow
\left[
\begin{array}{cc}
e^{i\sigy(\xi^L_{+} - \frac{\phi_+}{2})}
&0\\
0&
e^{i\sigy(\xi^L_{-} - \frac{\phi_-}{2})}
\end{array}
\right]
\left[
\begin{array}{cc}
e^{i\mu_+\sigz}
&0\\
0&
e^{i\mu_-\sigz}
\end{array}
\right]
\left[
\begin{array}{cc}
e^{i\sigy(\xi^R_{+} + \frac{\phi_+}{2})}
&0\\
0&
e^{i\sigy(\xi^R_{-} + \frac{\phi_-}{2})}
\end{array}
\right]
\udarrow
\;,
\label{eq-tilv-3mat-prod}
\eeq
where

\beq
e^{i\mu_\pm\sigz}
=
p_{\pm} + i \sigz q_{\pm}
\;.
\eeq

Next we want to
calculate the effect of
a similarity transformation
$\unor \udarrow(\cdot)\udarrow\unor^\dagger$
on each of the 3 matrices
that are being multiplied on the
right hand side of Eq.(\ref{eq-tilv-3mat-prod}).
For any real numbers $\alpha, \beta$,

\begin{subequations}\label{eq-m-ud-1}
\begin{eqnarray}
\lefteqn{
\unor
\udarrow
\left[
\begin{array}{cc}
e^{i\alpha\sigy}&0\\
0&e^{i\beta\sigy}
\end{array}
\right]
\udarrow
\unor^\dagger
=}\nonumber\\
&=&
\unor
\udarrow
e^{i\alpha\sigy(0)\nbar(1)}
e^{i\beta\sigy(0)n(1)}
\udarrow
\unor^\dagger\\
&=&
\unor
\udarrow
e^{i\frac{\alpha}{2}(\siguy + \sigzy)}
e^{i\frac{\beta}{2}(\siguy - \sigzy)}
\udarrow
\unor^\dagger\\
&=&
e^{i(\frac{-\alpha-\beta}{2})\sigyu}
e^{i(\frac{ \alpha-\beta}{2})\siguy}
\;,
\end{eqnarray}
\end{subequations}
and

\begin{subequations}\label{eq-m-ud-2}
\begin{eqnarray}
\lefteqn{
\unor
\udarrow
\left[
\begin{array}{cc}
e^{i\alpha\sigz}&0\\
0&e^{i\beta\sigz}
\end{array}
\right]
\udarrow
\unor^\dagger
=}\nonumber\\
&=&
\unor
\udarrow
e^{i\frac{\alpha}{2}(\siguz + \sigzz)}
e^{i\frac{\beta}{2}(\siguz - \sigzz)}
\udarrow
\unor^\dagger\\
&=&
e^{i(\frac{ \alpha+\beta}{2})\sigxx}
e^{i(\frac{ \alpha-\beta}{2})\sigzz}
\;.
\end{eqnarray}
\end{subequations}
Applying the identities given by
Eqs.(\ref{eq-m-ud-1}) and
(\ref{eq-m-ud-2}) to Eq.(\ref{eq-tilv-3mat-prod})
yields

\begin{subequations}\label{eq-u-pre-fin}
\begin{eqnarray}
U&=&
e^{i\frac{\theta_L}{2}\siguz}
\unor \tilde{V}  \unor^\dagger
e^{i\frac{\theta_R}{2}\siguz}
\\
&=&
e^{i\frac{\theta_L}{2}\siguz}
e^{\frac{i}{2}(\gamma_L\siguy + \gamma'_L\sigyu)}
e^{i(\mu'\sigxx + \mu\sigzz)}
e^{\frac{i}{2}(\gamma_R\siguy + \gamma'_R\sigyu)}
e^{i\frac{\theta_R}{2}\siguz}
\;.
\end{eqnarray}
\end{subequations}
A simple consequence of
the table given in Eq.(\ref{tab-uutoz})
is that:

\beq
e^{i(\mu'\sigxx + \mu\sigzz)}
=
\begin{array}{c}
\Qcircuit @C=1em @R=.5em @!R{
&\timesgate
&\gate{e^{i\mu\sigz}}
&\timesgate
&\qw
\\
&\dotgate\qwx[-1]
&\gate{e^{i\mu'\sigx}}
&\dotgate\qwx[-1]
&\qw
}
\end{array}
\;.
\label{eq-cnot-sigs-cnto}
\eeq
Eqs.(\ref{eq-u-pre-fin}) and (\ref{eq-cnot-sigs-cnto})
imply Eq.(\ref{eq-defla-u-out}).
\qed

Our next goal is to generalize
the above theorem. But first, let us make
some observations that will
pave the way towards this goal.

As is well known and easily proven,
for any $A\in SU(2)$, one can find real numbers
$\alpha, \beta$ and $\gamma$, such that:

\beq
A= e^{i\alpha\sigz}e^{i\beta\sigy}e^{i\gamma\sigz}
\;.
\eeq
A translation of this  last equation
into circuit language is:

\beq
\begin{array}{c}
\Qcircuit @C=1em @R=1em @!R{
&\emptygate
&\qw
}
\end{array}
=
\begin{array}{c}
\Qcircuit @C=1em @R=1em @!R{
&\gate{R_Z}
&\gate{R_Y}
&\gate{R_Z}
&\qw
}
\end{array}
\;.
\label{eq-a-2-rz-ry-rz}
\eeq

Any $A\in SU(2)$ can be diagonalized
by a unitary matrix $U$:
\beq
A = U e^{i\theta\sigz} U^\dagger
\;.
\eeq
Therefore,

\beq
[A(0)]^{n(1)}=
U(0) [e^{i\theta\sigz(0)}]^{n(1)} U(0)^\dagger
\;.
\eeq
A translation of this  last equation
into circuit language is:

\beq
\begin{array}{c}
\Qcircuit @C=1em @R=1em @!R{
&\emptygate
&\qw
\\
&\dotgate\qwx[-1]
&\qw
}
\end{array}
=
\begin{array}{c}
\Qcircuit @C=1em @R=.5em @!R{
&\emptygate
&\gate{R_Z}
&\emptygate
&\qw
\\
&\qw
&\dotgate\qwx[-1]
&\qw
&\qw
}
\end{array}
\;.
\label{eq-rz-new}
\eeq

Since $n=\frac{1}{2}(1-\sigz) $,

\beqa
e^{i\theta \sigz(1)n(0)}
&=&
e^{i\frac{\theta}{2} \sigz(1)[1-\sigz(0)]}\\
&=&
e^{i\frac{\theta}{2}\{
\sigz(1)
-\sigz(0)
+\sigz(0)[1-\sigz(1)]
\}}
\\
&=&
e^{i\frac{\theta}{2}[\sigz(1)
-\sigz(0)]}
e^{i\theta\sigz(0)n(1)}
\;.
\eeqa
A translation of this  last equation
into circuit language is:

\beq
\begin{array}{c}
\Qcircuit @C=1em @R=.5em @!R{
&\dotgate\qwx[1]
&\qw
\\
&\gate{R_Z}
&\qw
}
\end{array}
=
\begin{array}{c}
\Qcircuit @C=1em @R=.5em @!R{
&\gate{R_Z}
&\qw
&\gate{R_Z}
&\qw
\\
&\gate{R_Z}
&\qw
&\dotgate\qwx[-1]
&\qw
}
\end{array}
\;.
\label{eq-rz-flip}
\eeq

We are now ready to prove a
simple generalization of the Deflation Identity:

\begin{theo}\label{th-defla2}

\beq
\begin{array}{c}
\Qcircuit @C=1em @R=1em @!R{
&\emptygate
&\emptygate
&\emptygate
&\qw
\\
&\dotgate\qwx[-1]
&\emptygate
&\dotgate\qwx[-1]
&\qw
}
\end{array}
=
\begin{array}{c}
\Qcircuit @C=1em @R=1em @!R{
&\emptygate
&\timesgate
&\emptygate
&\timesgate
&\emptygate
&\qw
\\
&\emptygate
&\dotgate\qwx[-1]
&\emptygate
&\dotgate\qwx[-1]
&\emptygate
&\qw
}
\end{array}
\;,
\label{eq-deflat2-same-side}
\eeq
and

\beq
\begin{array}{c}
\Qcircuit @C=1em @R=1em @!R{
&\emptygate
&\emptygate
&\dotgate
&\qw
\\
&\dotgate\qwx[-1]
&\emptygate
&\emptygate\qwx[-1]
&\qw
}
\end{array}
=
\begin{array}{c}
\Qcircuit @C=1em @R=1em @!R{
&\emptygate
&\timesgate
&\emptygate
&\timesgate
&\emptygate
&\qw
\\
&\emptygate
&\dotgate\qwx[-1]
&\emptygate
&\dotgate\qwx[-1]
&\emptygate
&\qw
}
\end{array}
\;.
\label{eq-deflat2-oppo-side}
\eeq
\end{theo}
\proof

By virtue of Eqs.(\ref{eq-a-2-rz-ry-rz})
and (\ref{eq-rz-new}),
\beqa
\begin{array}{c}
\Qcircuit @C=1em @R=1em @!R{
&\emptygate
&\emptygate
&\emptygate
&\qw
\\
&\dotgate\qwx[-1]
&\emptygate
&\dotgate\qwx[-1]
&\qw
}
\end{array}
&=&
\begin{array}{c}
\Qcircuit @C=1em @R=1em @!R{
&\emptygate
&\gate{R_Z}
&\emptygate
&\gate{R_Z}
&\emptygate
&\qw
\\
&\qw
&\dotgate\qwx[-1]
&\emptygate
&\dotgate\qwx[-1]
&\qw
&\qw
}
\end{array}\\
&=&
\begin{array}{c}
\Qcircuit @C=1em @R=1em @!R{
&\emptygate
&\gate{R_Z}
&\gate{R_Y}
&\gate{R_Z}
&\emptygate
&\qw
\\
&\emptygate
&\dotgate\qwx[-1]
&\gate{R_Y}
&\dotgate\qwx[-1]
&\emptygate
&\qw
}
\end{array}
\;.
\eeqa
Applying the Deflation Identity to
the right hand side of
the last equation establishes
Eq.(\ref{eq-deflat2-same-side}).

By virtue of Eqs.(\ref{eq-rz-new})
and (\ref{eq-rz-flip}),
\beq
\begin{array}{c}
\Qcircuit @C=1em @R=1em @!R{
&\emptygate
&\emptygate
&\dotgate
&\qw
\\
&\dotgate\qwx[-1]
&\emptygate
&\emptygate\qwx[-1]
&\qw
}
\end{array}
=
\begin{array}{c}
\Qcircuit @C=1em @R=1em @!R{
&\emptygate
&\emptygate
&\gate{R_Z}
&\qw
&\qw
\\
&\dotgate\qwx[-1]
&\emptygate
&\dotgate\qwx[-1]
&\emptygate
&\qw
}
\end{array}
\;.
\eeq
Applying Eq.(\ref{eq-deflat2-same-side}) to
the right hand side of
the last equation establishes
Eq.(\ref{eq-deflat2-oppo-side}).
\qed

\section{Two Applications of
Deflation Identity}
In this section, we will
give two application of the
Deflation Identity,
one to ``opening
and closing a breach" in a quantum
circuit, the other to
the CS decomposition of
a 2-qubit operator.

\subsection{Opening and Closing a Breach}
\begin{quote}
{\it
Once more unto the breach, dear friends, once more;
    Or close the wall up with our English dead!}
(from ``King Henry V"
by W. Shakespeare)
\end{quote}

\begin{theo}\label{th-closing-breach}
(Closing a breach)
If
\beq
U=
\begin{array}{c}
\Qcircuit @C=1em @R=1em @!R{
&\timesgate
&\gate{B}
&\timesgate
&\gate{A}
&\timesgate
&\qw
\\
&\dotgate\qwx[-1]
&\emptygate
&\dotgate\qwx[-1]
&\qw
&\dotgate\qwx[-1]
&\qw
}
\end{array}
\;,
\eeq
then

\beq
U=
\begin{array}{c}
\Qcircuit @C=1em @R=1em @!R{
&\emptygate
&\timesgate
&\emptygate
&\timesgate
&\emptygate
&\qw
\\
&\emptygate
&\dotgate\qwx[-1]
&\emptygate
&\dotgate\qwx[-1]
&\emptygate
&\qw
}
\end{array}
\;.
\eeq
\end{theo}
\proof

\beqa
U&=&
\begin{array}{c}
\Qcircuit @C=1em @R=1em @!R{
&\timesgate
&\gate{BA}
&\gate{A^\dagger}
&\timesgate
&\gate{A}
&\timesgate
&\qw
\\
&\dotgate\qwx[-1]
&\emptygate
&\qw
&\dotgate\qwx[-1]
&\qw
&\dotgate\qwx[-1]
&\qw
}
\end{array}\\
&=&
\begin{array}{c}
\Qcircuit @C=1em @R=1em @!R{
&\timesgate
&\gate{BA}
&\gate{A^\dagger \sigx A\sigx}
&\qw
\\
&\dotgate\qwx[-1]
&\emptygate
&\dotgate\qwx[-1]
&\qw
}
\end{array}
\;.
\eeqa
An application of the Deflation Identity
to the right hand side of the last equation
establishes the theorem.
\qed

The last theorem shows how
one can ``close a breach" within a quantum circuit,
while simultaneously reducing the
number of CNOTs in the circuit by 1.
It is also possible to ``open a breach"
within a circuit;
i.e., replace the circuit
by an equivalent one that
has a breach:

\beq
\begin{array}{c}
\Qcircuit @C=1em @R=1em @!R{
&\multigate{1}{T}
&\emptygate
&\emptygate
&\emptygate
&\multigate{1}{U}
&\qw
\\
&\ghost{T}
&\dotgate\qwx[-1]
&\emptygate
&\dotgate\qwx[-1]
&\ghost{U}
&\qw
}
\end{array}
=
\begin{array}{c}
\Qcircuit @C=1em @R=1em @!R{
&\multigate{1}{T'}
&\emptygate
&\emptygate
&\emptygate
&\multigate{1}{U'}
&\qw
\\
&\ghost{T}
&\dotgate\qwx[-1]
&\qw
&\dotgate\qwx[-1]
&\ghost{U}
&\qw
}
\end{array}
\;.
\eeq
In the above circuit,
we assume that the total
number of CNOTs within the sub-circuits
$U$ and $T$
does not increase
when they are replaced
by $U'$ and $T'$.
A useful strategy for
reducing the number of CNOTs
in a circuit by 1 is to open a breach
within the circuit
without increasing
its number of CNOTs, and then to
close the breach using
Theorem \ref{th-closing-breach}.
In a future paper, we
will say more about this strategy for
reducing the number of CNOTs
in a circuit. We will show that the
strategy also works
for N-qubit circuits with $N>2$.

\subsection{Expressing 2-qubit CSD as a circuit with 3 CNOTs}

According to the CS decomposition\cite{Golub}\cite{Rudi},
any 2-qubit unitary operation $U$
can be expressed as:

\beq
U =
e^{i\alpha}
\left[
\begin{array}{cc}
e^{i\alpha_L}L_0 & 0\\
0 & e^{-i\alpha_L}L_1
\end{array}
\right]
\left[
\begin{array}{cc}
C & S \\
-S & C
\end{array}
\right]
\left[
\begin{array}{cc}
e^{i\alpha_R}R_0 & 0\\
0 & e^{-i\alpha_R}R_1
\end{array}
\right]
\;,
\label{eq-gen-2bit-csd}
\eeq
where

\beq
C=diag(\cos\theta_1, \cos\theta_2)
\;,\;\;
S=diag(\sin\theta_1, \sin\theta_2)
\;,
\eeq
and
$\alpha, \alpha_L,
\alpha_R, \theta_1, \theta_2\in \RR$, and
$L_0, L_1, R_0, R_1\in SU(2)$.
Our goal for this section is to express
the right hand side of
Eq.(\ref{eq-gen-2bit-csd})
as a circuit with just 3 CNOTs
(and some 1-qubit rotations).
Ref.\cite{VD} showed how,
given any 2-qubit unitary operation
$U$, one can first perform a
KAK1 decomposition\cite{Tuc-KAK}
of $U$, and then
express the outcome as
a 3 CNOT circuit. What we give below is
an alternative method
for expressing a 2-qubit unitary operation
as a quantum circuit with 3 CNOTs. Our method
is via the CS decomposition,
rather than via KAK1.

Note that given any two $2\times 2$ matrices $A, B$,
\beq
\left[
\begin{array}{cc}
A & 0\\
0 & B
\end{array}
\right]
=
A(0)^{\nbar(1)}
B(0)^{n(1)}
=
A(0)
[A(0)^\dagger B(0)]^{n(1)}
\;.
\eeq
A translation of this  last equation
into circuit language is:

\beq
\left[
\begin{array}{cc}
A & 0\\
0 & B
\end{array}
\right]
=
\begin{array}{c}
\Qcircuit @C=1em @R=1em @!R{
&\gate{A}
&\gate{B}
&\qw
\\
&\ogate\qwx[-1]
&\dotgate\qwx[-1]
&\qw
}
\end{array}
=
\begin{array}{c}
\Qcircuit @C=1em @R=1em @!R{
&\gate{A}
&\gate{A^\dagger B}
&\qw
\\
&\qw
&\dotgate\qwx[-1]
&\qw
}
\end{array}
\;.
\eeq
Therefore,

\beq
\left[
\begin{array}{cc}
e^{i\alpha_L}L_0 & 0\\
0 & e^{-i\alpha_L}L_1
\end{array}
\right]
=
e^{i\alpha_L\sigz(1)}
\left[
\begin{array}{cc}
L_0 & 0\\
0 & L_1
\end{array}
\right]
=
\begin{array}{c}
\Qcircuit @C=1em @R=1em @!R{
&\gate{L_0}
&\gate{L_0^\dagger L_1}
&\qw
\\
&\gate{e^{i\alpha_L\sigz}}
&\dotgate\qwx[-1]
&\qw
}
\end{array}
\;.
\label{eq-csd-piece1}
\eeq
Note also that

\beq
\left[
\begin{array}{cc}
C & S\\
-S & C
\end{array}
\right]
=
\udarrow
\left[
\begin{array}{cc}
e^{i\theta_1\sigy} & 0\\
0 & e^{i\theta_2\sigy}
\end{array}
\right]
\udarrow
=
\begin{array}{c}
\Qcircuit @C=1em @R=1em @!R{
&\ogate
&\dotgate
&\qw
\\
&\gate{e^{i\theta_1\sigy}}\qwx[-1]
&\gate{e^{i\theta_2\sigy}}\qwx[-1]
&\qw
}
\end{array}
\;.
\label{eq-csd-piece2}
\eeq
By virtue of Eqs.(\ref{eq-csd-piece1}) and
(\ref{eq-csd-piece2}),
we can express the
CS Decomposition
Eq.(\ref{eq-gen-2bit-csd}) in circuit language
as follows:

\beq
U=
e^{i\alpha}
\begin{array}{c}
\Qcircuit @C=1em @R=1em @!R{
&\gate{L_0}
&\gate{L_0^\dagger L_1}
&\ogate
&\dotgate
&\gate{R_1 R_0^\dagger}
&\gate{R_0}
&\qw
\\
&\gate{e^{i\alpha_L\sigz}}
&\dotgate\qwx[-1]
&\gate{e^{i\theta_1\sigy}}\qwx[-1]
&\gate{e^{i\theta_2\sigy}}\qwx[-1]
&\dotgate\qwx[-1]
&\gate{e^{i\alpha_R\sigz}}
&\qw
}
\end{array}
\;.
\label{eq-csd-ckt1}
\eeq

Note that
\beq
\begin{array}{c}
\Qcircuit @C=1em @R=1em @!R{
&\ogate
&\dotgate
&\qw
\\
&\gate{e^{i\theta_1\sigy}}\qwx[-1]
&\gate{e^{i\theta_2\sigy}}\qwx[-1]
&\qw
}
\end{array}
=
\begin{array}{c}
\Qcircuit @C=1em @R=1em @!R{
&\qw
&\dotgate
&\qw
&\dotgate
&\qw
\\
&\gate{e^{i\frac{\theta_1 + \theta_2}{2}\sigy }}
&\dotgate\qwx[-1]
&\gate{e^{i \frac{\theta_1 - \theta_2}{2}\sigy}}
&\dotgate\qwx[-1]
&\qw
}
\end{array}
\;,
\label{eq-sy-sy}
\eeq
a result which is easily proven
by considering the two possible cases $n(0)=0,1$.
Note that
\beq
\begin{array}{c}
\Qcircuit @C=1em @R=1em @!R{
&\dotgate
&\gate{R_1 R_0^\dagger}
&\qw
\\
&\dotgate\qwx[-1]
&\dotgate\qwx[-1]
&\qw
}
\end{array}
=
\begin{array}{c}
\Qcircuit @C=1em @R=1em @!R{
&\gate{\sigz R_1 R_0^\dagger}
&\qw
\\
&\dotgate\qwx[-1]
&\qw
}
\end{array}
\;.
\label{eq-absorb-cnot}
\eeq
We can always find 2-dimensional unitary matrices
$U_L$ and $U_R$ such that:

\beq
L^\dagger_0 L_1=
U_L
e^{i\lambda_L\sigz}
U_L^\dagger
\;,\;\;
\sigz R_1 R^\dagger_0
=
U_R
e^{i\lambda_R\sigz}
U_R^\dagger
\;.
\label{eq-ur-ul}
\eeq
Applying Eqs.(\ref{eq-sy-sy}),
(\ref{eq-absorb-cnot})
and (\ref{eq-ur-ul}) to Eq.(\ref{eq-csd-ckt1})
yields

\beq
U=
e^{i\alpha}
\begin{array}{c}
\Qcircuit @C=1em @R=1em @!R{
&\gate{L_0 U_L}
&\gate{e^{i\lambda_L\sigz}}
&\gate{U_L^\dagger}
&\dotgate
&\gate{U_R}
&\gate{e^{i\lambda_R\sigz}}
&\gate{U_R^\dagger R_0 }
\\
&\gate{e^{i\alpha_L\sigz}}
&\dotgate\qwx[-1]
&\gate{e^{i\frac{\theta_1 + \theta_2}{2}\sigy }}
&\dotgate\qwx[-1]
&\gate{e^{i\frac{\theta_1 - \theta_2}{2}\sigy }}
&\dotgate\qwx[-1]
&\gate{e^{i\alpha_R\sigz }}
}
\end{array}
\;.
\label{eq-last-explicit-ckt}
\eeq

Note that

\beq
[\sigz(0)]^{n(1)}
=
(-i)^{n(1)}[i\sigz(0)]^{n(1)}
=
e^{-i\frac{\pi}{4}}
e^{i\frac{\pi}{4}\sigz(1)}
e^{i\frac{\pi}{2}\sigz(0)n(1)}
\;.
\eeq
A translation of this  last equation
into circuit language is:

\beq
\begin{array}{c}
\Qcircuit @C=1em @R=1em @!R{
&\dotgate
&\qw
\\
&\dotgate\qwx[-1]
&\qw
}
\end{array}
=
\begin{array}{c}
\Qcircuit @C=1em @R=1em @!R{
&\qw
&\gate{i\sigz}
&\qw
\\
&\gate{(-i)^n}
&\dotgate\qwx[-1]
&\qw
}
\end{array}
=
e^{-i\frac{\pi}{4}}
\begin{array}{c}
\Qcircuit @C=1em @R=1em @!R{
&\qw
&\gate{e^{i\frac{\pi}{2}\sigz}}
&\qw
\\
&\gate{e^{i\frac{\pi}{4}\sigz}}
&\dotgate\qwx[-1]
&\qw
}
\end{array}
\;.
\label{eq-not-not}
\eeq
Eq.(\ref{eq-not-not}) and
the Deflation Identity,
together, imply that

\beq
\begin{array}{c}
\Qcircuit @C=1em @R=1em @!R{
&\dotgate
&\gate{R_Y}
&\gate{R_Z}
&\qw
\\
&\dotgate\qwx[-1]
&\gate{R_Y}
&\dotgate\qwx[-1]
&\qw
}
\end{array}
=
e^{-i\frac{\pi}{4}}
\begin{array}{c}
\Qcircuit @C=1em @R=1em @!R{
&\emptygate
&\dotgate
&\emptygate
&\dotgate
&\emptygate
&\qw
\\
&\emptygate
&\dotgate\qwx[-1]
&\emptygate
&\dotgate\qwx[-1]
&\emptygate
&\qw
}
\end{array}
\;.
\label{eq-two-not-not}
\eeq
We can reduce the right hand side of
Eq.(\ref{eq-last-explicit-ckt}) to a
quantum circuit containing just 3 CNOTs
by applying the identity
Eq.(\ref{eq-two-not-not}) twice:

\beqa
U &=& e^{i\alpha}
\begin{array}{c}
\Qcircuit @C=1em @R=1em @!R{
&\emptygate
&\gate{R_Z}
&\gate{R_Y}
&\gate{R_Z}
&\dotgate
&\gate{R_Y}
&\gate{R_Z}
&\emptygate
&\qw
\\
&\emptygate
&\dotgate\qwx[-1]
&\gate{R_Y}
&\qw
&\dotgate\qwx[-1]
&\gate{R_Y}
&\dotgate\qwx[-1]
&\emptygate
&\qw
}
\end{array}
\\
&=&e^{i(\alpha -\frac{\pi}{4})}
\begin{array}{c}
\Qcircuit @C=1em @R=1em @!R{
&\emptygate
&\gate{R_Z}
&\emptygate
&\dotgate
&\emptygate
&\dotgate
&\emptygate
&\qw
\\
&\emptygate
&\dotgate\qwx[-1]
&\emptygate
&\dotgate\qwx[-1]
&\emptygate
&\dotgate\qwx[-1]
&\emptygate
&\qw
}
\end{array}
\\
&=&
e^{i(\alpha -\frac{\pi}{4})}
\begin{array}{c}
\Qcircuit @C=1em @R=1em @!R{
&\emptygate
&\gate{R_Z}
&\gate{R_Y}
&\dotgate
&\emptygate
&\dotgate
&\emptygate
&\qw
\\
&\emptygate
&\dotgate\qwx[-1]
&\gate{R_Y}
&\dotgate\qwx[-1]
&\emptygate
&\dotgate\qwx[-1]
&\emptygate
&\qw
}
\end{array}
\\
&=&
e^{i\alpha}
\begin{array}{c}
\Qcircuit @C=1em @R=1em @!R{
&\emptygate
&\dotgate
&\emptygate
&\dotgate
&\emptygate
&\dotgate
&\emptygate
&\qw
\\
&\emptygate
&\dotgate\qwx[-1]
&\emptygate
&\dotgate\qwx[-1]
&\emptygate
&\dotgate\qwx[-1]
&\emptygate
&\qw
}
\end{array}
\;.
\eeqa

\end{document}